\documentclass[12pt,a4paper]{article}
\pdfoutput=1
\usepackage[utf8]{inputenc}
\usepackage[T1]{fontenc}
\usepackage[english]{babel}
\usepackage{amsmath, amssymb, amsthm}
\usepackage{float}
\usepackage{geometry}
\usepackage{xcolor}
\usepackage{hyperref}
\usepackage{caption}
\usepackage{tikz}
\usetikzlibrary{3d,arrows.meta,calc,decorations.markings,positioning,intersections}

\hypersetup{
    colorlinks=true,
    linkcolor=blue,
    urlcolor=blue,
    pdftitle={Tight-Binding Spectra of Finite Incidence Geometries: From Spatial Localization to $SU(6)$ Flavor Symmetry}
}
\title{\textbf{Tight-Binding Spectra of Finite Incidence Geometries: From Spatial Localization to $SU(6)$ Flavor Symmetry}}
\author{
    \textbf{Paweł Nurowski} \\
    \small Centrum Fizyki Teoretycznej, Polska Akademia Nauk, Al. Lotników 32/46, 02-668 Warszawa, Poland \\
    \small Guangdong Technion Israel Institute of Technology, No. 241, Daxue Road, Shantou, China \\
    \small \texttt{nurowski@cft.edu.pl}
}
\date{\today}
\begin{document}

\maketitle

\begin{abstract}
  We investigate the spectral properties of tight-binding quantum Hamiltonians defined on the bipartite Levi graphs of finite geometric configurations. In Part I, we analyze the $10_3$ Desargues and Kantor configurations, demonstrating how deterministic spatial deformations induced by real planar embeddings ($\mathbb{R}\mathbb{P}^2$) destroy translational symmetry, leading to structural wave localization. We show how embedding in the complex projective plane ($\mathbb{C}\mathbb{P}^2$) restores Bloch wave propagation via synthetic gauge phases. In Part II, we evaluate the Schläfli double six ($12_5, 30_2$) and the Cremona-Richmond ($15_3$) configurations in $\mathbb{R}^3$. We analytically derive their exact spectral decompositions, confirming the existence of macroscopic zero-energy flat bands ($E=0$) driven entirely by geometric frustration. In Part III, we establish a formal structural isomorphism between these discrete tight-binding networks and the flavor symmetry sector of the Standard Model. We map the Schläfli graph to the $SU(6)$ flavor multiplets, where the flat band corresponds to the kinematic freezing of ultra-heavy baryons. Finally, we discuss the complementary Cremona-Richmond $15_3$ topology, demonstrating how its distinct geometric nature (based on tritangent planes rather than point intersections) provides a purely algebraic, topological completion to the $W(E_6)$ symmetry of the 27 lines.
\end{abstract}

\section{Introduction}
Tight-binding Hamiltonians operating on discrete lattice graphs provide a fundamental framework for modeling quantum dynamics, conventionally applied to electrons propagating through physical crystalline solids. In this work, we generalize this approach by replacing standard physical lattices with the incidence graphs (Levi graphs) of highly symmetric, historic algebraic geometries. 

This paper is structured to illustrate a conceptual transition from dynamics in physical space to transformations in abstract quantum state spaces. In Part I, we treat the vertices of the $10_3$ configurations as physical spatial coordinates. We analyze the conditions for unhindered Bloch wave propagation and demonstrate how geometric constraints during spatial embedding act as deterministic scattering potentials, inducing wave localization. 

In Parts II and III, we shift our paradigm from physical space to abstract Fock space. By evaluating the Schläfli double six and the Cremona-Richmond configurations, we treat the network vertices not as spatial coordinates, but as distinct multiparticle flavor states. Under this interpretation, the hopping amplitudes represent structural transformations between fundamental fields (quarks) and bound states (mesons). Through exact spectral decomposition, we reveal how the purely mathematical phenomena of geometric frustration and flat-band localization on these finite graphs structurally mirror the kinematic confinement and multiplet structures of the $SU(6)$ flavor symmetry of the Standard Model.

\vspace{0.5cm}

\section*{PART I: Localization and Wave Propagation in $10_3$ Configurations}
\addcontentsline{toc}{section}{PART I: Localization and Wave Propagation in $10_3$ Configurations}

\section{Geometric Configurations: Desargues vs. Kantor}
We consider two topologically distinct structures from the class of $10_3$ geometric configurations.
The first is the Desargues configuration \cite{desargues, Coxeter1950}, which exhibits a highly symmetric, bipartite incidence structure but lacks cyclic translational symmetry.
The second is Kantor's cyclic configuration \cite{Kantor1881}, characterized by an intrinsic $\mathbb{Z}_{10}$ cyclic permutation symmetry.
Each configuration generates a bipartite incidence graph (Levi graph) $\mathcal{G} = (\mathcal{V}, \mathcal{E})$ consisting of $|\mathcal{V}|
= 20$ vertices, partitioned into a set of 10 points $\mathcal{P}$ and 10 lines $\mathcal{L}$.
Let $\mathbb{K}$ denote the scalar field $\mathbb{R}$ or $\mathbb{C}$. The associated Hilbert space $\mathcal{H} \cong \mathbb{K}^{20}$ is defined by the orthonormal basis vectors spanning the points and lines: $\mathcal{H} = \text{span}(\{|p_i\rangle\}_{i=0}^9, \{|l_j\rangle\}_{j=0}^9)$.
Under this algebraic formulation, lines can be represented as 3-forms in the exterior algebra space $\bigwedge^3 \mathbb{K}^{10}$: $|l_j\rangle = |p_a \wedge p_b \wedge p_c\rangle$.
\definecolor{levired}{RGB}{220, 50, 50}
\definecolor{leviblue}{RGB}{50, 100, 220}
\definecolor{levigold}{RGB}{218, 165, 32}

\begin{figure}[H]
\centering
\begin{minipage}{0.48\textwidth}
\centering
\begin{tikzpicture}[scale=0.4, transform shape]
    \def\R{5.5} 
    \def\N{20}  
    \def\step{360/\N}

    \foreach \i in {0,...,9} {
        \pgfmathsetmacro{\angleP}{90 - (2*\i)*\step}
        \node[circle, draw=blue!80, fill=white, thick, minimum size=0.7cm, inner sep=0pt, font=\small] (p\i) at (\angleP:\R) {$p_{\i}$};
        \pgfmathsetmacro{\angleL}{90 - (2*\i + 1)*\step}
        \node[circle, draw=red!80, fill=white, thick, minimum size=0.7cm, inner sep=0pt, font=\small] (l\i) at (\angleL:\R) {$l_{\i}$};
    }

    \foreach \i in {0,...,9} {
        \pgfmathsetmacro{\nextP}{int(mod(\i+1,10))}
        \draw[levired, thick] (p\i) -- (l\i);
        \draw[leviblue, thick] (l\i) -- (p\nextP);
    }

    \foreach \i in {0,2,4,6,8} {
        \pgfmathsetmacro{\chordTarget}{int(mod(\i+2,10))}
        \draw[levigold, thick] (p\i) -- (l\chordTarget);
    }
    \foreach \i in {1,3,5,7,9} {
        \pgfmathsetmacro{\chordTarget}{int(mod(\i+4,10))}
        \draw[levigold, thick] (p\i) -- (l\chordTarget);
    }
    \node at (0.5, 3) {\large \textbf{Desargues}};
\end{tikzpicture}
\end{minipage}\hfill
\begin{minipage}{0.48\textwidth}
\centering
\begin{tikzpicture}[scale=0.4, transform shape]
    \def\R{5.5} 
    \def\N{20}  
    \def\step{360/\N}

    \foreach \i in {0,...,9} {
        \pgfmathsetmacro{\angleP}{90 - (2*\i)*\step}
        \node[circle, draw=blue!80, fill=white, thick, minimum size=0.7cm, inner sep=0pt, font=\small] (p\i) at (\angleP:\R) {$p_{\i}$};
        \pgfmathsetmacro{\angleL}{90 - (2*\i + 1)*\step}
        \node[circle, draw=red!80, fill=white, thick, minimum size=0.7cm, inner sep=0pt, font=\small] (l\i) at (\angleL:\R) {$l_{\i}$};
    }

    \foreach \i in {0,...,9} {
        \pgfmathsetmacro{\nextP}{int(mod(\i+1,10))}
        \pgfmathsetmacro{\chordTarget}{int(mod(\i+3,10))}
        \draw[levired, thick] (p\i) -- (l\i);
        \draw[leviblue, thick] (l\i) -- (p\nextP);
        \draw[levigold, thick] (l\i) -- (p\chordTarget);
    }
    \node at (0, 0) {\large \textbf{Kantor}};
\end{tikzpicture}
\end{minipage}
\caption{\small Geometric representation of the 20-vertex Levi graphs for the evaluated $10_3$ configurations.
\textbf{Left:} The Desargues configuration graph, exhibiting internal chord structures that form a closed resonant cavity.
\textbf{Right:} Kantor's cyclic configuration graph, where the rotationally symmetric internal chords establish a periodic 1D network.}
\label{fig:levi_comparison}
\end{figure}
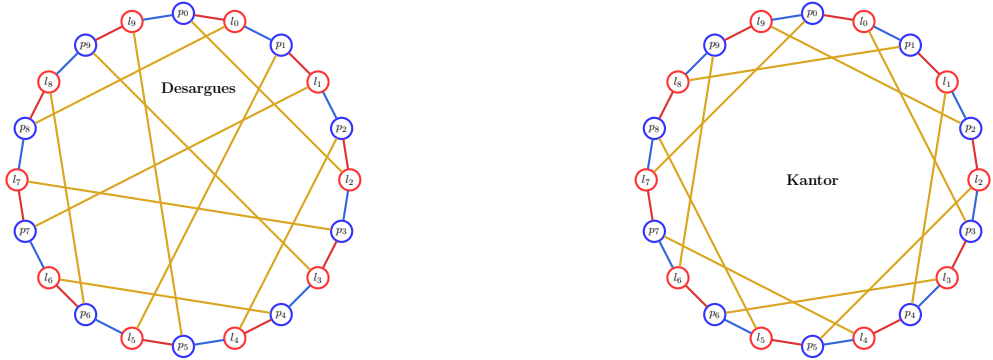
\section{The Quantum Network Hamiltonian}
We construct a tight-binding Hamiltonian based on the incidence relation $\mathcal{I}$:
$$\hat{H} = -t \sum_{(p_i, l_j) \in \mathcal{I}} \left( |p_i\rangle\langle l_j| + |l_j\rangle\langle p_i| \right)$$
where $t > 0$ is the hopping amplitude between an incident point and line.

\subsection{The Desargues Configuration (No Bloch Waves)}
For the Desargues configuration~\cite{desargues, Coxeter1950}, the Hamiltonian spectrum is strictly integer:
$$ E_{Desargues} \in \{ -3t, -2t, -t, +t, +2t, +3t \} $$
There are no Bloch waves here. The eigenstates $|\Psi_E\rangle$ are static, quantized standing waves. The Desargues graph acts as a discrete closed system---a resonant cavity where quantum states are confined in symmetric oscillations.

To find the coefficients $c_{\{a,b\}}$, we map our abstract indices to the isomorphism between the Desargues configuration and the complete graph $K_5$---a regular combinatorial graph consisting of 5 vertices where every distinct pair of vertices is connected by a unique edge, yielding exactly $\binom{5}{2}=10$ edges. We establish a direct bijection between the 10 points of the Desargues configuration and the 10 distinct edges (pairs) of this complete graph $K_5$. We establish a bijection between the 10 points and the 10 distinct unordered pairs drawn from a 5-element set $S = \{1, 2, 3, 4, 5\}$. Thus, we define the point basis vectors as $|p_{\{a,b\}}\rangle$. Consequently, the 10 lines naturally map to the 10 triplets: $|l_{\{a,b,c\}}\rangle$. Incidence is strictly defined: a point lies on a line if and only if its representative pair is a subset of the triplet ($\{a,b\} \subset \{a,b,c\}$).

Because the Levi graph is bipartite, the eigenvalue equation possesses chiral symmetry. Let $|\phi\rangle_{\mathcal{P}}$ denote a 10-dimensional spatial base mode strictly localized within the point sub-lattice. For $E \neq 0$, the explicit full 20-dimensional eigenstate $|\Psi_{\pm E}\rangle \in \mathcal{H}$ is constructed by generating the corresponding line amplitudes via the Hamiltonian action:
$$ |\Psi_{\pm E}\rangle = \frac{1}{\mathcal{N}} \left( |\phi\rangle_{\mathcal{P}} \mp \frac{t}{|E|} \sum_{1 \le a < b < c \le 5} \Big( \langle p_{\{a,b\}} | \phi \rangle_{\mathcal{P}} + \langle p_{\{b,c\}} | \phi \rangle_{\mathcal{P}} + \langle p_{\{a,c\}} | \phi \rangle_{\mathcal{P}} \Big) |l_{\{a,b,c\}}\rangle \right) $$

Under this mapping, the 10 spatial point modes $|\psi^{(n)}\rangle \in \mathcal{H}_{\mathcal{P}}$ serve as the core structural generators for the total 20-dimensional eigenstates $|\Psi_{\pm E}\rangle \in \mathcal{H}$. To guide the reader through the explicit vector configurations listed below, the full states across all three physical excitation families are reconstructed from the small point modes $|\psi^{(n)}\rangle$ according to the following exact projection rules:

\begin{itemize}
    \item \textbf{The Isotropic Singlet State ($E = \pm 3t$, for $n = 1$):}
    $$ |\Psi_{\pm 3t}^{(1)}\rangle = \frac{1}{\sqrt{20}} \left( |\psi^{(1)}\rangle \mp \sum_{1 \le a < b < c \le 5} |l_{\{a,b,c\}}\rangle \right) $$
    
    \item \textbf{The Vector Multiplet States ($E = \pm 2t$, for $n = 2, 3, 4, 5$):}
    $$ |\Psi_{\pm 2t}^{(n)}\rangle = \frac{1}{\mathcal{N}_v} \left( |\psi^{(n)}\rangle \mp \frac{1}{2} \sum_{1 \le a < b < c \le 5} \Big( \langle p_{\{a,b\}} | \psi^{(n)}\rangle + \langle p_{\{b,c\}} | \psi^{(n)}\rangle + \langle p_{\{a,c\}} | \psi^{(n)}\rangle \Big) |l_{\{a,b,c\}}\rangle \right) $$
    
    \item \textbf{The Tensor Multiplet States ($E = \pm 1t$, for $n = 6, 7, 8, 9, 10$):}
    $$ |\Psi_{\pm 1t}^{(n)}\rangle = \frac{1}{\mathcal{N}_t} \left( |\psi^{(n)}\rangle \mp \sum_{1 \le a < b < c \le 5} \Big( \langle p_{\{a,b\}} | \psi^{(n)}\rangle + \langle p_{\{b,c\}} | \psi^{(n)}\rangle + \langle p_{\{a,c\}} | \psi^{(n)}\rangle \Big) |l_{\{a,b,c\}}\rangle \right) $$
\end{itemize}
where $\mathcal{N}_v$ and $\mathcal{N}_t$ denote the exact quantum normalization parameters. Guided by these definitions, the explicit spanning basis of the spatial point generators decomposes into the following three families governed by the irreducible representations of the $S_5$ symmetry group:

\paragraph{1. The Isotropic Singlets ($E = \pm 3t$, Multiplicity: 1)}
Representing the trivial representation of $S_5$. The single spatial base state $|\psi^{(1)}\rangle$ is the totally symmetric coherent superposition of all 10 geometric points. The sum of amplitudes mapped to any line $\{a,b,c\}$ is exactly $1+1+1=3$. With $|E/t| = 3$, the prefactor yields $\mp 1$. The explicit state vector is:
\begin{align*}
    |\psi^{(1)}\rangle &= |p_{\{1,2\}}\rangle + |p_{\{1,3\}}\rangle + |p_{\{1,4\}}\rangle + |p_{\{1,5\}}\rangle + |p_{\{2,3\}}\rangle \\
    &\quad + |p_{\{2,4\}}\rangle + |p_{\{2,5\}}\rangle + |p_{\{3,4\}}\rangle + |p_{\{3,5\}}\rangle + |p_{\{4,5\}}\rangle
\end{align*}

\paragraph{2. The Vector Excitations ($E = \pm 2t$, Multiplicity: 4)}
Corresponding to the 4-dimensional standard representation of $S_5$. These 4 spatial base modes act as primary dipole excitations. They are parameterized by a 5-dimensional weight vector $\vec{v}$ satisfying $\sum_{k=1}^5 v_k = 0$, where the amplitude on any point is $c_{\{a,b\}} = v_a + v_b$. The explicit spanning basis of state vectors is generated by selecting four independent weight assignments:
\begin{align*}
    |\psi^{(2)}\rangle &\leftarrow \vec{v} = (+1, -1, 0, 0, 0) \implies \\
    &|\psi^{(2)}\rangle = |p_{\{1,3\}}\rangle + |p_{\{1,4\}}\rangle + |p_{\{1,5\}}\rangle - |p_{\{2,3\}}\rangle - |p_{\{2,4\}}\rangle - |p_{\{2,5\}}\rangle \\
    |\psi^{(3)}\rangle &\leftarrow \vec{v} = (0, +1, -1, 0, 0) \implies \\
    &|\psi^{(3)}\rangle = |p_{\{1,2\}}\rangle + |p_{\{2,4\}}\rangle + |p_{\{2,5\}}\rangle - |p_{\{1,3\}}\rangle - |p_{\{3,4\}}\rangle - |p_{\{3,5\}}\rangle \\
    |\psi^{(4)}\rangle &\leftarrow \vec{v} = (0, 0, +1, -1, 0) \implies \\
    &|\psi^{(4)}\rangle = |p_{\{1,3\}}\rangle + |p_{\{2,3\}}\rangle + |p_{\{3,5\}}\rangle - |p_{\{1,4\}}\rangle - |p_{\{2,4\}}\rangle - |p_{\{4,5\}}\rangle \\
    |\psi^{(5)}\rangle &\leftarrow \vec{v} = (0, 0, 0, +1, -1) \implies \\
    &|\psi^{(5)}\rangle = |p_{\{1,4\}}\rangle + |p_{\{2,4\}}\rangle + |p_{\{3,4\}}\rangle - |p_{\{1,5\}}\rangle - |p_{\{2,5\}}\rangle - |p_{\{3,5\}}\rangle 
\end{align*}

\paragraph{3. The Tensor Excitations ($E = \pm 1t$, Multiplicity: 5)}
Corresponding to the 5-dimensional irreducible representation of $S_5$. These 5 spatial base modes form higher-order multipole standing waves. They are constructed by defining superpositions on fundamental 5-cycles (pentagons) within $K_5$ with alternating $+1$ weights on the cycle edges and $-1$ on the complement (the pentagram), ensuring total amplitude cancellation around specific local structures. 

To form a mathematically valid basis, the chosen pentagons must be linearly independent. The explicit spanning basis of 5 strictly independent state vectors is generated by selecting the following fundamental cycles:
{\small \begin{align*}
    |\psi^{(6)}\rangle &= |p_{\{1,2\}}\rangle + |p_{\{2,3\}}\rangle + |p_{\{3,4\}}\rangle + |p_{\{4,5\}}\rangle + |p_{\{1,5\}}\rangle - |p_{\{1,3\}}\rangle - |p_{\{1,4\}}\rangle - |p_{\{2,4\}}\rangle - |p_{\{2,5\}}\rangle - |p_{\{3,5\}}\rangle \\
    |\psi^{(7)}\rangle &= |p_{\{1,2\}}\rangle + |p_{\{2,3\}}\rangle + |p_{\{3,5\}}\rangle + |p_{\{4,5\}}\rangle + |p_{\{1,4\}}\rangle - |p_{\{1,3\}}\rangle - |p_{\{1,5\}}\rangle - |p_{\{2,4\}}\rangle - |p_{\{2,5\}}\rangle - |p_{\{3,4\}}\rangle \\
    |\psi^{(8)}\rangle &= |p_{\{1,2\}}\rangle + |p_{\{2,4\}}\rangle + |p_{\{3,4\}}\rangle + |p_{\{3,5\}}\rangle + |p_{\{1,5\}}\rangle - |p_{\{1,3\}}\rangle - |p_{\{1,4\}}\rangle - |p_{\{2,3\}}\rangle - |p_{\{2,5\}}\rangle - |p_{\{4,5\}}\rangle \\
    |\psi^{(9)}\rangle &= |p_{\{1,2\}}\rangle + |p_{\{2,4\}}\rangle + |p_{\{4,5\}}\rangle + |p_{\{3,5\}}\rangle + |p_{\{1,3\}}\rangle - |p_{\{1,4\}}\rangle - |p_{\{1,5\}}\rangle - |p_{\{2,3\}}\rangle - |p_{\{2,5\}}\rangle - |p_{\{3,4\}}\rangle \\
    |\psi^{(10)}\rangle &= |p_{\{1,2\}}\rangle + |p_{\{2,5\}}\rangle + |p_{\{3,5\}}\rangle + |p_{\{3,4\}}\rangle + |p_{\{1,4\}}\rangle - |p_{\{1,3\}}\rangle - |p_{\{1,5\}}\rangle - |p_{\{2,3\}}\rangle - |p_{\{2,4\}}\rangle - |p_{\{4,5\}}\rangle
\end{align*}}
These 10 base vectors $|\psi^{(1)}\rangle \dots |\psi^{(10)}\rangle$ mathematically span the exact 10-dimensional symmetric subspace of the spatial point modes, providing the complete generator basis for the full 20-dimensional spectrum.

\subsection{Kantor's Cyclic Graph (Emergence of Bloch Waves)}
Kantor's cyclic configuration acts as a 1D periodic lattice~\cite{Kantor1881}.
Here, the $\mathbb{Z}_{10}$ translational symmetry allows us to apply Bloch's theorem~\cite{Bloch1928}.
We demonstrate how the geometry dictates the wave propagation.
Because the Levi graph is bipartite, the Hamiltonian acts locally on the points and lines.
According to the incidence rules of Kantor's cyclic graph, a point $p_n$ is connected to the line $l_n$, the adjacent line $l_{n-1}$, and the chord line $l_{n-3}$ (modulo 10).
Thus, the Hamiltonian acts on a point state as:
$$ \hat{H} |p_n\rangle = -t \left( |l_n\rangle + |l_{n-1}\rangle + |l_{n-3}\rangle \right) $$
By symmetry, the action on a line state shifts the indices in the opposite direction:
$$ \hat{H} |l_n\rangle = -t \left( |p_n\rangle + |p_{n+1}\rangle + |p_{n+3}\rangle \right) $$

To find the stationary states, we construct a Bloch wave ansatz with a quasimomentum $k \in \{0, 1, \dots, 9\}$.
Let $\phi = \frac{2\pi}{10}$. We assign unknown internal amplitudes $A_k$ to the points and $B_k$ to the lines:
$$ |\psi_k\rangle = \sum_{n=0}^{9} e^{i k n \phi} \left( A_k |p_n\rangle + B_k |l_n\rangle \right) $$

Applying the Schr\"{o}dinger equation $\hat{H} |\psi_k\rangle = E_k |\psi_k\rangle$ to the point sub-lattice, the hopping terms accumulate structural phase shifts. Factoring out the shifts yields:
$$ \hat{H} \sum_{n=0}^{9} e^{i k n \phi} A_k |p_n\rangle = -t A_k \sum_{n=0}^{9} e^{i k n \phi} \left( 1 + e^{i k \phi} + e^{i 3 k \phi} \right) |l_n\rangle $$
We can encapsulate the geometric steps into a single complex structure factor $f_k = 
1 + e^{i k \phi} + e^{i 3 k \phi}$. Similarly, acting on the lines yields the complex conjugate $f_k^*$.

This reduces the 20-dimensional graph into a $2 \times 2$ local eigenvalue problem:
\begin{align*}
    -t f_k^* B_k &= E_k A_k \\
    -t f_k A_k &= E_k B_k
\end{align*}
Multiplying these two equations reveals the energy dispersion. Since $f_k f_k^* = |f_k|^2$, we get:
$$ E_k = \pm t |f_k|
= \pm t \left| 1 + e^{i \frac{2\pi k}{10}} + e^{i \frac{3 \cdot 2\pi k}{10}} \right|
$$
This yields the exact energy band of our quantum network model.
To find the eigenvectors, we express the complex structure factor in polar form: $f_k = |f_k| e^{i \theta_k}$.
Substituting this back into the eigenvalue equations yields $B_k = \mp e^{i \theta_k} A_k$.
Setting $A_k = 1$ and adding the normalization factor $\frac{1}{\sqrt{20}}$, we arrive at the exact analytical form of the propagating \textbf{Bloch waves}:
$$|\psi_k^\pm\rangle = \frac{1}{\sqrt{20}} \sum_{n=0}^{9} e^{i \frac{2\pi k n}{10}} \left( |p_n\rangle \pm e^{i \theta_k} |l_n\rangle \right)$$

In its mathematical form, the phase $\theta_k$ maps the local chord geometry into the wave's internal degrees of freedom.
Furthermore, from a condensed matter perspective, the winding of this internal phase $\theta_k$ as $k$ traverses the 1D Brillouin zone defines a global topological invariant known as the Zak phase~\cite{Zak1989}.
This non-trivial geometric phase acts as a topological protector of the propagating states.

\section{Planar Embedding: Breaking the Translational Symmetry}
In 1881, Kantor represented the configurations on a flat plane ($\mathbb{R}^2$) using straight lines~\cite{Kantor1881}.
Eight years later, Schr\"{o}ter proved that the cyclic $10_3$ configuration \textbf{cannot} be drawn in $\mathbb{R}\mathbb{P}^2$ with straight lines~\cite{Schroeter1889}. The constraint of using straight lines breaks the $\mathbb{Z}_{10}$ translation symmetry.

\subsection{Deterministic Structural Localization}
In the Hamiltonian, this geometric constraint introduces deterministic spatial deformations $\delta t_{ij}$. The Bloch waves scatter off these structural defects. Through destructive interference, the waves collapse into tightly localized states. Unlike classical Anderson localization \cite{Anderson1958}, which relies on stochastic disorder, this geometric deformation deterministically disrupts the Bloch waves, transitioning the finite system from a periodic conductor into an insulator.

\subsection{Symmetry Restoration in $\mathbb{C}\mathbb{P}^2$}
Embedding the graph over the field of real numbers breaks its cyclic symmetry. However, in the complex projective plane ($\mathbb{C}\mathbb{P}^2$), Kantor's cyclic configuration can be embedded using straight lines. By utilizing quantum phases (where a geometric bend is offset by a complex gauge phase $t \rightarrow t e^{i\phi_{ij}}$), the cyclic symmetry is mathematically restored:
$$\hat{H}_{\mathbb{C}} = -t \sum \left( e^{i\phi_{ij}} |p_i\rangle\langle l_j| + e^{-i\phi_{ij}} |l_j\rangle\langle p_i| \right)$$
The Bloch waves are restored. They propagate as chiral, topologically protected states, unconstrained by the limitations of the flat $\mathbb{R}^2$ space, effectively bypassing the geometric localization through synthetic gauge fields.

\vspace{1cm}

\section*{PART II: Spectral Analysis of the Schläfli Double Six}
\addcontentsline{toc}{section}{PART II: Spectral Analysis of the Schläfli Double Six}

\section{Incidence Structure and Graph Representation}
While the configurations in Part I represented networks embedded in physical space where hopping amplitudes denote spatial movement, the Levi graph of the Schläfli double six configuration is fundamentally different. Here, we analyze it as an abstract state space (a Fock space), where the vertices represent multiparticle states and the hopping amplitudes denote structural transformations between these internal configurations.

The Schläfli double six configuration \cite{Schlafli1858}, denoted as $(12_5, 30_2)$, consists of 12 lines divided into two distinct disjoint families, $\mathcal{L}_A = \{l_1 \dots l_6\}$ and $\mathcal{L}_B = \{L_1 \dots L_6\}$, which generate 30 intersection points.

The incidence relation requires that a line $l_i \in \mathcal{L}_A$ intersects a line $L_j \in \mathcal{L}_B$ if and only if $i \neq j$.
Each intersecting pair defines a unique point $p_{ij}$, yielding $6 \times 5 = 30$ point states.
The associated bipartite Levi graph comprises 42 vertices and 60 edges, with its global automorphism group governed by the non-abelian direct product $S_6 \times \mathbb{Z}_2$.
The absence of an underlying translation group implies that Bloch wave solutions are strictly forbidden;
instead, the geometry behaves as a multi-dimensional bounded resonant cavity.

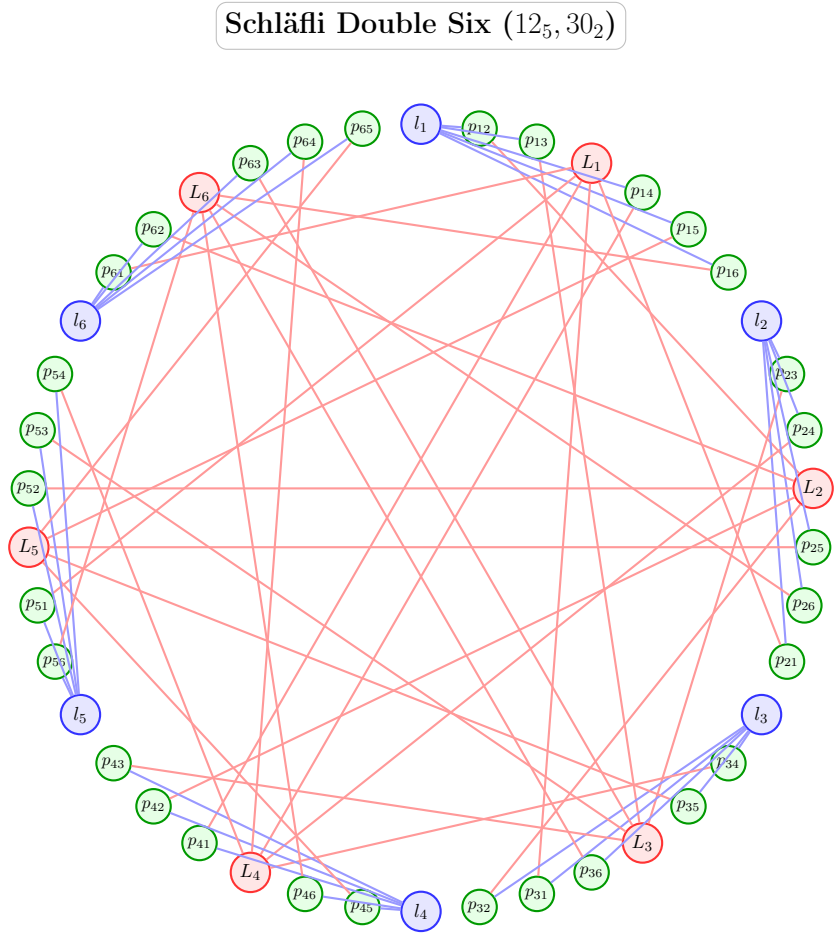
\begin{figure}[H]
\centering
\begin{tikzpicture}[scale=0.65, transform shape]
    \def\R{8} 
    \def\N{42}
    \def\step{360/\N}

    \node[fill=white!90, inner sep=5pt, draw=gray!50, rounded corners] at (0,10) {\Large \textbf{Schläfli Double Six ($12_5, 30_2$)}};
    \foreach \i/\k in {1/0, 2/7, 3/14, 4/21, 5/28, 6/35} {
        \pgfmathsetmacro{\ang}{90 - \k*\step}
        \node[circle, draw=blue!80, fill=blue!10, thick, minimum size=0.8cm, inner sep=0pt, font=\small] (l\i) at (\ang:\R) {$l_{\i}$};
    }

    \foreach \j/\k in {1/3, 2/10, 3/17, 4/24, 5/31, 6/38} {
        \pgfmathsetmacro{\ang}{90 - \k*\step}
        \node[circle, draw=red!80, fill=red!10, thick, minimum size=0.8cm, inner sep=0pt, font=\small] (L\j) at (\ang:\R) {$L_{\j}$};
    }

    \foreach \i/\j/\k in {
        1/2/1, 1/3/2, 1/4/4, 1/5/5, 1/6/6,
        2/3/8, 2/4/9, 2/5/11, 2/6/12, 2/1/13,
        3/4/15, 3/5/16, 3/6/18, 3/1/19, 3/2/20,
        4/5/22, 4/6/23, 4/1/25, 4/2/26, 4/3/27,
        5/6/29, 5/1/30, 5/2/32, 5/3/33, 5/4/34,
        6/1/36, 6/2/37, 6/3/39, 6/4/40, 6/5/41
    } {
        
        \pgfmathsetmacro{\ang}{90 - \k*\step}
        \node[circle, draw=green!60!black, fill=green!10, thick, minimum size=0.7cm, inner sep=0pt, font=\footnotesize] (P\i\j) at (\ang:\R) {$p_{\i\j}$};
    }

    \foreach \i in {1,...,6} {
        \foreach \j in {1,...,6} {
            \ifnum\i=\j\else
                \draw[blue!40, thick] (l\i) -- (P\i\j);
                \draw[red!40, thick] (L\j) -- (P\i\j);
            \fi
        }
    }
\end{tikzpicture}
\caption{\small The 42-node Levi Graph of the Schläfli Double Six.
The notation explicitly reflects the underlying geometry: the first family of lines is denoted by $l_i$ (blue), the second family of lines by $L_j$ (red), and their incidence is represented by the intersection points $p_{ij}$ (green).
A true Hamiltonian cycle is mathematically forbidden by the unequal bipartite nature ($12 \neq 30$), but this symmetrical layout highlights the $S_6$ rotational geometry governing the flat band.}
\label{fig:schlafli_single_circle}
\end{figure}

The corresponding bipartite Levi graph has 42 vertices and 60 edges.
Unlike the cyclic Kantor graph, its symmetry is governed by the highly non-abelian group $S_6 \times \mathbb{Z}_2$.
Consequently, \textbf{Bloch waves are strictly forbidden}. Instead, this geometry acts as a multidimensional resonant cavity.

\section{Spectral Analysis and Geometric Quantization}
To uncover the physics of the double six, we rigorously define the Hamiltonian dynamics on its Levi graph. The total Hilbert space is $42$-dimensional, constructed as a direct sum of the line subspace and the point subspace: $\mathcal{H} = \mathcal{H}_{\mathcal{L}} \oplus \mathcal{H}_{\mathcal{P}} \cong \mathbb{C}^{12} \oplus \mathbb{C}^{30}$. 

We define the tight-binding Hamiltonian operator $\hat{H}$ that couples these spaces according to the strict incidence rules of the double six:
$$ 
\begin{aligned} 
\hat{H} = -t \sum_{i=1}^6 \sum_{\substack{j=1 \\ j \neq i}}^6 \Big( & |l_i\rangle\langle p_{ij}| + |L_j\rangle\langle p_{ij}| \\
+ & |p_{ij}\rangle\langle l_i| + |p_{ij}\rangle\langle L_j| \Big) 
\end{aligned} 
$$

Let an arbitrary state in this total space be written with amplitudes $A_i$ for the first family of lines, $B_j$ for the second family, and $C_{ij}$ for the intersection points:
$$ |\Psi\rangle = \sum_{i=1}^6 A_i |l_i\rangle + \sum_{j=1}^6 B_j |L_j\rangle + \sum_{i \neq j} C_{ij} |p_{ij}\rangle $$

Applying the Schrödinger equation $\hat{H}|\Psi\rangle = E|\Psi\rangle$ yields a system of coupled constraints for the amplitudes. Projecting the equation onto the basis vectors, we get:
\begin{align}
    \langle p_{ij} | \hat{H} |\Psi\rangle &= -t (A_i + B_j) = E C_{ij} \label{eq:points} \\
    \langle l_i | \hat{H} |\Psi\rangle &= -t \sum_{j \neq i} C_{ij} = E A_i \label{eq:lines_a} \\
    \langle L_j | \hat{H} |\Psi\rangle &= -t \sum_{i \neq j} C_{ij} = E B_j \label{eq:lines_b}
\end{align}

\subsection{Subspace Reduction and the Effective Matrix}
Because the graph is bipartite, chiral symmetry guarantees that non-zero energy states come in pairs $\pm E$. For $E \neq 0$, we can explicitly integrate out the point sub-lattice. By substituting the geometric constraint from Eq.~(\ref{eq:points}) directly into Eq.~(\ref{eq:lines_a}) and multiplying by $E$, we obtain the effective action of the squared Hamiltonian solely on the line amplitudes $A_i$:
$$ E^2 A_i = -t \sum_{j \neq i} E C_{ij} = -t \sum_{j \neq i} \left[ -t (A_i + B_j) \right] = t^2 \sum_{j \neq i} (A_i + B_j) $$
Since each line contains exactly 5 points, the sum over $A_i$ yields $5 A_i$. Thus, the equation simplifies to:
$$ E^2 A_i = 5t^2 A_i + t^2 \sum_{j \neq i} B_j $$
By symmetry, the exact same operation for $B_j$ yields:
$$ E^2 B_j = 5t^2 B_j + t^2 \sum_{i \neq j} A_i $$

This mathematically proves that within the 12-dimensional subspace of lines $\mathcal{H}_{\mathcal{L}}$, the squared Hamiltonian $\hat{H}^2$ reduces to a block matrix form:
$$ \hat{H}^2_{\mathcal{L}} = t^2 \begin{pmatrix} 5 \mathbb{I}_6 & \textbf{J}_6 - \mathbb{I}_6 \\ \textbf{J}_6 - \mathbb{I}_6 & 5 \mathbb{I}_6 \end{pmatrix} $$
where $\mathbb{I}_6$ is the identity matrix and $\textbf{J}_6$ is the all-ones matrix.

\subsection{Explicit Base Vectors and Multiplets}
Diagonalizing this reduced matrix allows us to explicitly construct the base eigenvectors. Following the irreducible representations of the $S_6$ symmetry group, the $42$-dimensional space decomposes into exactly four discrete, highly degenerate multiplets. 

We construct the explicit base vectors for each multiplet below:

\paragraph{1. The Isotropic Singlets ($E = \pm \sqrt{10}t$, Multiplicity: 2)}
These represent the totally symmetric (trivial) representation of $S_6$. The amplitudes are uniform across all lines: $A_i = 1$ and $B_j = 1$. From Eq.~(\ref{eq:points}), the amplitude on the points is $C_{ij} = \mp \frac{2t}{\sqrt{10}t}$. The explicit full eigenstates are:
$$ |\Psi_{\pm \sqrt{10}t}\rangle \propto \sum_{i=1}^6 |l_i\rangle + \sum_{j=1}^6 |L_j\rangle \mp \sqrt{\frac{2}{5}} \sum_{i \neq j} |p_{ij}\rangle $$

\paragraph{2. The Tensor Multiplet ($E = \pm 2t$, Multiplicity: 10)}
These 10 states are constructed using a 6-dimensional weight vector $\vec{v}$ whose elements sum to zero ($\sum_{k=1}^6 v_k = 0$). By assigning symmetric amplitudes $A_i = v_i$ and $B_j = v_j$, the off-diagonal sum evaluates to $\sum_{j \neq i} v_j = -v_i$. The effective squared energy is $E^2 v_i = 5t^2 v_i + t^2 (-v_i) = 4t^2 v_i$, yielding $E = \pm 2t$. 
The base spatial modes spanning this subspace are defined explicitly by 5 independent traceless vectors (e.g., $\vec{v}^{(1)} = (1, -1, 0, 0, 0, 0)$, etc.), giving the states:
$$ |\Psi_{\pm 2t}^{(\vec{v})}\rangle \propto \sum_{i=1}^6 v_i |l_i\rangle + \sum_{j=1}^6 v_j |L_j\rangle \mp \frac{1}{2} \sum_{i \neq j} (v_i + v_j) |p_{ij}\rangle $$

\paragraph{3. The Vector Multiplet ($E = \pm \sqrt{6}t$, Multiplicity: 10)}
Using the same traceless weight vector $\vec{v}$, we now assign antisymmetric amplitudes: $A_i = v_i$ and $B_j = -v_j$. The effective squared energy becomes $E^2 v_i = 5t^2 v_i + t^2 \sum_{j \neq i} (-v_j) = 6t^2 v_i$, yielding $E = \pm \sqrt{6}t$. This generates the vector multiplet:
$$ |\Psi_{\pm \sqrt{6}t}^{(\vec{v})}\rangle \propto \sum_{i=1}^6 v_i |l_i\rangle - \sum_{j=1}^6 v_j |L_j\rangle \mp \frac{1}{\sqrt{6}} \sum_{i \neq j} (v_i - v_j) |p_{ij}\rangle $$

\paragraph{4. The Flat Band ($E = 0$, Multiplicity: 20)}
A prominent feature of the Schläfli geometry is its highly degenerate zero-energy null space. For $E=0$, Eqs.~(\ref{eq:lines_a}) and (\ref{eq:lines_b}) dictate that the amplitudes on all 12 lines must strictly vanish: $A_i = B_j = 0$. The wave is entirely confined to the 30 intersection points. 
To satisfy $\hat{H}|\Psi\rangle = 0$, the sum of point amplitudes along any line must be zero. We explicitly construct the 20 basis vectors spanning this flat band by selecting fundamental 4-cycles (rectangles) within the bipartite geometry. Choosing two distinct lines from family A ($i, k$) and two from family B ($j, m$), the localized basis mode is:
$$ |\Psi_{E=0}^{(i,j,k,m)}\rangle = |p_{ij}\rangle - |p_{kj}\rangle + |p_{km}\rangle - |p_{im}\rangle $$
When the quantum wave attempts to propagate from point $p_{ij}$ onto line $l_i$, it is met with the exact opposite phase $-|p_{im}\rangle$ arriving from the same structural loop. The geometry enforces total destructive interference without any spatial disorder. The probability amplitude is strictly localized. The particle is trapped.

Unlike Kantor's graph, which required the complex plane, Schläfli proved that the double six exists without geometric distortion within 27 straight lines upon a cubic surface in standard three-dimensional Euclidean space ($\mathbb{R}^3$). Thus, $\mathbb{R}^3$ is finally sufficient to host this geometric flat band insulation mechanism physically.

\vspace{1cm}

\section*{PART IIa: Spectral Analysis of the Cremona-Richmond Configuration}
\addcontentsline{toc}{section}{PART IIa: Spectral Analysis of the Cremona-Richmond Configuration}

\section{Incidence Structure and the $15_3$ Levi Graph}
The algebraic classification of lines upon a general cubic surface in $\mathbb{R}^3$ establishes the existence of exactly 27 straight lines \cite{Schlafli1858, Cartan_27}. Isolating a Schläfli double six (12 lines) leaves a uniquely defined complementary set of exactly 15 lines. 

The incidence geometry of these 15 lines and their 15 associated tritangent planes (synthemes) forms the highly symmetric Cremona-Richmond configuration, denoted as a $15_3$ configuration. In this bipartite network, each line lies on exactly 3 tritangent planes, and each plane contains exactly 3 lines (forming a triangle upon that plane). 

Algebraically, this configuration is isomorphic to the generalized quadrangle $GQ(2,2)$. The 15 tritangent planes $\Pi = \{\pi_1 \dots \pi_{15}\}$ possess a natural bijection to the 15 unordered pairs $\{a,b\}$ drawn from a 6-element set. The 15 lines $\mathcal{L} = \{l_1 \dots l_{15}\}$ map to the 15 synthemes (the partitions of the 6 elements into three mutually disjoint pairs). Incidence is strictly defined: a plane contains a line if and only if the plane's representative pair is a constituent of the line's syntheme.

The associated bipartite Levi graph comprises 30 vertices and 45 edges. This graph is uniquely identifiable as the Tutte 8-cage. Its global automorphism group is isomorphic to the symmetric group $S_6$.

\section{Algebraic Representation and Exact Spectral Decomposition}
The total 30-dimensional Hilbert space is $\mathcal{H} = \mathcal{H}_{\Pi} \oplus \mathcal{H}_{\mathcal{L}}$. The tight-binding Hamiltonian operator $\hat{H}$ couples the planes and lines uniformly with hopping amplitude $t$:
$$ \hat{H} = -t \sum_{(\pi_{\{a,b\}}, l_j) \in \mathcal{I}} \left( |\pi_{\{a,b\}}\rangle\langle l_j| + |l_j\rangle\langle \pi_{\{a,b\}}| \right) $$

For non-zero energy eigenvalues ($E \neq 0$), the explicit full 30-dimensional eigenstates $|\Psi_{\pm E}\rangle$ are constructed by mapping the plane amplitudes to the lines. The requirement that this mapping is an eigenfunction reduces the plane amplitudes to the eigenvalue problem of the Kneser graph $KG(6, 2)$, dictating three highly degenerate multiplets:

\paragraph{1. The Isotropic Singlets ($E = \pm 3t$, Multiplicity: 2)}
Representing the trivial representation of $S_6$. The spatial plane amplitude is uniformly $1$.
$$ |\Psi_{\pm 3t}\rangle = \frac{1}{\sqrt{30}} \left( \sum_{1 \le a < b \le 6} |\pi_{\{a,b\}}\rangle \mp \sum_{j=1}^{15} |l_j\rangle \right) $$

\paragraph{2. The Tensor Multiplet ($E = \pm 2t$, Multiplicity: 18)}
Corresponding to the 9-dimensional irreducible representation of $S_6$. We parameterize these states using a $6 \times 6$ symmetric matrix $\mathbf{W}$ with a zero diagonal ($W_{aa} = 0$), which satisfies the strict row-sum constraint $\sum_{b=1}^6 W_{ab} = 0$ for every row $a$. Imposing these 6 independent linear constraints onto the 15 off-diagonal variables mathematically reduces the parameter space to exactly 9 dimensions.

$$ |\Psi_{\pm 2t}^{(\textbf{W})}\rangle \propto \sum_{1 \le a < b \le 6} W_{ab} |\pi_{\{a,b\}}\rangle \mp \frac{1}{2} \sum_{l_j \in \mathcal{L}} \Big( W_{a_j b_j} + W_{c_j d_j} + W_{e_j f_j} \Big) |l_j\rangle $$

\paragraph{3. The Geometric Flat Band ($E = 0$, Multiplicity: 10)}
Because the graph is bipartite, the 10-dimensional null space strictly decouples: exactly 5 eigenstates are localized completely on the tritangent planes, and 5 are localized completely on the lines. Parameterized by a traceless vector $\vec{v}$, the plane-localized states are:
$$ |\Psi_{E=0, \Pi}^{(\vec{v})}\rangle = \frac{1}{\sqrt{20}} \sum_{1 \le a < b \le 6} (v_a + v_b) |\pi_{\{a,b\}}\rangle $$
The geometry forces total destructive interference at every line node.

\section*{PART III: Structural Isomorphisms to the Standard Model}
\addcontentsline{toc}{section}{PART III: Structural Isomorphisms to the Standard Model}

\section{The $SU(6)$ Flavor Symmetry of the Schläfli Graph}
The $S_6 \times \mathbb{Z}_2$ automorphism group governing the discrete tight-binding model on the Schläfli double six configuration provides a formal structural isomorphism to the flavor symmetry sector of the Standard Model of particle physics. If we consider the six known quark flavors ($u, d, s, c, b, t$), their approximate flavor symmetry is described by the Lie group $SU(6)$, whose Weyl permutation subgroup is isomorphic to the symmetric group $S_6$. The discrete $\mathbb{Z}_2$ factor corresponds directly to the charge conjugation operation ($C$-symmetry), mapping matter to antimatter states. 

To establish the algebraic exactness of this isomorphism, we map the geometric network components to fundamental fields, and the corresponding Hamiltonian eigenstates to physical particle multiplets.

\subsection{Network Nodes as Fundamental Fields and Mesons}
The 42 vertices of the Schläfli Levi graph establish a static algebraic framework:
\begin{itemize}
    \item \textbf{The 12 Lines:} Map directly to the fundamental fermion degrees of freedom. The six lines of family $\mathcal{L}_A$ correspond to the \textbf{6 quark flavors} ($u, d, s, c, b, t$), while the six lines of family $\mathcal{L}_B$ correspond to the \textbf{6 antiquark flavors} ($\bar{u}, \bar{d}, \bar{s}, \bar{c}, \bar{b}, \bar{t}$).
    \item \textbf{The 30 Intersection Points:} A point $p_{ij}$ exists only at the intersection of line $l_i$ and line $L_j$ for $i \neq j$. In the particle model, this maps to the bound state of a quark $q_i$ and an antiquark $\bar{q}_j$ of a different flavor, defining the \textbf{30 off-diagonal mesons} within the $SU(6)$ framework (e.g., $u\bar{s} \equiv K^+$, $c\bar{d} \equiv D^+$). The geometry enumerates all valid off-diagonal flavor combinations.
\end{itemize}

\subsection{Global Eigenstates as Baryons and Resonance Multiplicities}
The global standing wave solutions over the configuration graph correspond to the multiplet structures found in hadronic physics:
\begin{itemize}
    \item \textbf{The 20 Flat-Band States ($E = 0$):} Combinatorially, selecting 3 distinct quark flavors from a set of 6 options yields $\binom{6}{3} = 20$ configurations. In the tight-binding model, these 20 states are strictly confined to the point sub-lattice by destructive interference. In quantum chromodynamics, this corresponds to the 20-dimensional multiplet of completely antisymmetric \textbf{heavy baryons} (e.g., $uds$, $cbs$, $tcb$). The zero-energy null space of the geometric configuration is exactly isomorphic to the state-space of the heavy baryon multiplet.
    \item \textbf{The 20 Dynamic Resonances ($E = \pm \sqrt{6}t$ and $E = \pm 2t$):} The remaining 10 vector and 10 tensor states have non-zero eigenvalues ($E \neq 0$), allowing active amplitude transport between lines and points. In the flavor model, these correspond to \textbf{dynamic meson resonances} or higher-order exotic states (such as tetraquarks), representing excitations characterized by a continuous coupling between the valence quark fields and the meson cloud.
\end{itemize}

\section{Kinematic Freezing and the Hadronization Timescale}
The structural link between the Schläfli flat-band resonator and the heavy baryon multiplets offers a formal topological analog for the phenomenon of kinematic freezing. In the discrete geometric network, the 20 flat-band states exhibit zero group velocity ($v_g = \frac{\partial E}{\partial k} = 0$) and cannot propagate through space due to geometric frustration within localized hexagonal loops.

In high-energy particle physics, an analogous dynamical constraint manifests in ultra-heavy baryons containing top quarks (such as a $tcb$ baryon). The mass of the top quark is sufficiently high that its weak decay lifetime ($\tau \sim 5 \times 10^{-25}$ s) is significantly shorter than the strong interaction timescale ($\tau_{\text{had}} \sim \Lambda_{\text{QCD}}^{-1} \sim 10^{-23}$ s) required for complete color confinement and spatial hadronization. Consequently, just as quantum wave propagation is suppressed within the double six configuration due to destructive interference, these heavy baryons are kinematically frozen at their point of creation, unable to propagate through space as asymptotic states. Our tight-binding model corresponds to the infinite mass limit of the strong interaction, where these flat-band states are topologically stable ($E \in \mathbb{R}$). To fully capture their finite lifetime induced by the weak decay, the geometric Hamiltonian must be extended by a non-Hermitian decay operator ($-i\Gamma$), which shifts the flat band into the complex plane, mathematically bridging the static structural freezing with the physical particle decay.

\section{The 27 Lines and the Topological Limits of Flavor Mapping}
The mapping of the Schläfli configuration to the Standard Model flavor symmetry relies strictly on the physical interpretation of vertices: fundamental lines as fermions, and their two-line intersection points as bipartite mesons. However, the completion of this geometry to the full 27 lines of the cubic surface---governed by the $W(E_6)$ group---introduces the Cremona-Richmond configuration.

As demonstrated in Part IIa, the $15_3$ Cremona-Richmond Levi graph defines its incidence not through simple two-line intersections, but through tritangent planes (synthemes) containing three intersecting lines (forming a triangle). This higher-order topological structure breaks the simple bipartite physical analogy. A single node representing a tritangent plane (encompassing six disjoint flavor fields) does not correspond to an asymptotic meson state in the Standard Model.

Therefore, while the Schläfli $12_5$ configuration provides a direct, localized topological isomorphism to the $SU(6)$ hadronic multiplets, the complementary $15_3$ configuration represents a purely algebraic, non-local topological completion. The union of these two sub-networks mathematically reconstructs the global $W(E_6)$ symmetry of Grand Unification, but their physical interpretations remain strictly distinct: one governs local bound states (mesons and baryons), while the other governs the abstract geometry of the enclosing algebraic surface.

\section{Conclusions}
In this work, we evaluated tight-binding quantum Hamiltonians defined on the bipartite Levi graphs of finite algebraic geometries. The analysis of $10_3$ configurations demonstrated how planar embeddings break translational symmetry, inducing deterministic spatial localization.

By shifting the interpretation of vertices to abstract flavor states, we evaluated the Schläfli double six geometry in $\mathbb{R}^3$. The spectral decomposition of its 42-node graph revealed highly degenerate multiplets, including a 20-dimensional zero-energy flat band ($E=0$) driven by geometric frustration. We established a formal isomorphism mapping this Schläfli geometry to the $SU(6)$ flavor sector of the Standard Model, where the flat band topologically models the kinematic freezing of ultra-heavy baryons.

Finally, we analyzed the complementary $15_3$ Cremona-Richmond configuration (the Tutte 8-cage). We corrected the interpretation of its geometric nodes, proving that its spectrum emerges from the incidence of lines and tritangent planes. Consequently, we concluded that while the Schläfli network successfully maps to localized hadronic states, the Cremona-Richmond $15_3$ topology acts as an abstract, non-local algebraic completion to the $W(E_6)$ symmetry, establishing clear geometric boundaries for physical flavor isomorphisms.


\begin{thebibliography}{99}

\bibitem{Anderson1958}
Anderson, P. W., \emph{Absence of Diffusion in Certain Random Lattices}. Physical Review, \textbf{109}(5), 1492--1505 (1958).

\bibitem{Bloch1928}
Bloch, F., \emph{Ueber die Quantenmechanik der Elektronen in Kristallgittern}. Zeitschrift f\"{u}r Physik, \textbf{52}(7--8), 555--600 (1928).

\bibitem{Cartan_27}
Cartan, \'{E}., \emph{Sur la g\'{e}om\'{e}trie des 27 droites d'une surface cubique}. {\OE}uvres compl\`{e}tes, Partie III (G\'{e}om\'{e}trie, Cin\'{e}matique), Gauthier-Villars, Paris (1955).

\bibitem{Coxeter1950}
Coxeter, H. S. M., \emph{Self-dual configurations and regular graphs}. Bulletin of the American Mathematical Society, \textbf{56}(5), 413--455 (1950).

\bibitem{desargues}
Desargues, G., (1639) ``Brouillon Projet d'une atteinte aux \'{e}v\'{e}nements des rencontres du C\^{o}ne avec un Plan'' \href{http://www.bibnum.education.fr/mathematiques/geometrie/brouillon-projet-d-une-atteinte-aux-evenements-des-rencontres-du-cone-avec-u}{online}

\bibitem{Kantor1881}
Kantor, S., \emph{Ueber die Configurationen (3, 3) mit den Indices 8, 9 und ihren Zusammenhang mit den Curven dritter Ordnung}. Sitzungsberichte der Mathematisch-Naturwissenschaftlichen Classe der Kaiserlichen Akademie der Wissenschaften, Wien, \textbf{84}, 915--932 (1881).

\bibitem{Ozawa2019}
Ozawa, T., et al., \emph{Topological photonics}. Reviews of Modern Physics, \textbf{91}(1), 015006 (2019).

\bibitem{Rechtsman2013}
Rechtsman, M. C., et al., \emph{Photonic Floquet topological insulators}. Nature, \textbf{496}(7444), 196--200 (2013).

\bibitem{Schlafli1858}
Schl\"{a}fli, L., \emph{An Attempt to Determine the Twenty-Seven Lines upon a Surface of the Third Order, and to Divide Such Surfaces into Species in Reference to the Reality of the Lines upon Them}. The Quarterly Journal of Pure and Applied Mathematics, \textbf{2}, 110--120 (1858).

\bibitem{Schroeter1889}
Schr\"{o}ter, H., \emph{Ueber die Bildungsweise und geometrische Construction der Configurationen $10_3$}. Nachrichten von der K\"{o}nigl. Gesellschaft der Wissenschaften und der Georg-Augusts-Universit\"{a}t zu G\"{o}ttingen, 193--236 (1889).

\bibitem{Zak1989}
Zak, J., \emph{Berry's phase for energy bands in solids}. Physical Review Letters, \textbf{62}(23), 2747--2750 (1989).

\end{thebibliography}
\end{document}